\documentstyle[twocolumn,aps,epsfig]{revtex}
\newcommand{\be}{\begin{equation}}
\newcommand{\ee}{\end{equation}}
\newcommand{\bea}{\begin{eqnarray}}
\newcommand{\eea}{\end{eqnarray}}
\renewcommand{\d}{{{\rm d}}}

\newcommand{\lton}{\mathrel{\lower.9ex
                  \hbox{$\stackrel{\displaystyle <}{\sim}$}}}

\begin{document}
\title{Pion Interferometry at RHIC: Probing a thermalized Quark-Gluon-Plasma?}
\author{Sven~Soff$^1$, Steffen~A.~Bass$^{2,3}$ and Adrian~Dumitru$^4$}
\address{$^1$Physics Department, Brookhaven National Laboratory, PO Box 5000, 
Upton, NY11973, USA\\
$^2$Department of Physics, Duke University, Durham, NC 27708, USA\\
$^3$RIKEN BNL Research Center, Brookhaven National Laboratory, 
Upton, NY11973, USA\\
$^4$Physics Department, Columbia University, 538 West 120th Street,
New York, NY10027, USA} 
\date{\today}
\maketitle   
\begin{abstract}
We calculate the Gaussian radius parameters of the pion-emitting source
in high energy heavy ion collisions, assuming a first order
phase transition from a thermalized Quark-Gluon-Plasma (QGP) 
to a gas of hadrons. Such a model leads to
a very long-lived dissipative {\em hadronic} rescattering phase which
dominates the properties of the two-pion correlation functions. The
radii are found to depend only weakly on the thermalization time
$\tau_i$, the critical temperature $T_c$ (and thus the latent heat),
and the specific entropy of the QGP. The dissipative hadronic stage enforces
large variations of the pion emission times around the mean. Therefore,
the model calculations suggest a rapid increase of $R_{\rm out}/R_{\rm side}$ 
as a function of $K_T$ if a thermalized QGP were formed.
\end{abstract}
\pacs{PACS numbers: 25.75.-q, 25.75.Ld, 12.38.Mh, 24.10.Lx}
\narrowtext

\vspace*{-0.9cm}

Bose-Einstein correlations in multiparticle production
processes~\cite{Goldhaber} provide valuable information on the 
space-time dynamics of fundamental
interactions~\cite{Shuryak:1973kq}.
In particular, lattice QCD calculations
predict the occurrence of a phase
transition at high temperature, and
it is hoped that correlations of identical pions
produced in high energy collisions of heavy ions lead to a better
understanding of the properties of that phase
transition (for a review on QGP signatures, see \cite{Bass:1999vz}).
A first order phase transition
leads to a prolonged hadronization time as compared to a cross-over
or ideal hadron gas with no phase transition, and has been related to unusually
large Hanbury-Brown--Twiss (HBT) radii~\cite{pratt86,schlei,dirk1}.
The phase of coexisting hadrons and QGP reduces the ``explosivity'' of the
high-density matter before hadronization, extending the emission duration of 
pions~\cite{pratt86,schlei,dirk1}.
This phenomenon should then depend on the
hadronization (critical) temperature $T_c$ and the latent heat of the
transition. For recent reviews on this topic we refer
to~\cite{reviews,wiedemannrep}.

Here, we investigate if and how HBT radii, characterizing the
pion source in the final state when all strong interactions have ceased,
depend on the properties of the QGP and the hadronization temperature.
The QGP is modeled as an ideal fluid undergoing hydrodynamic expansion
with a bag model equation of state, eventually
hadronizing via a first order phase transition~\cite{dirk1,gersdorff}. 
For simplicity, cylindrically
symmetric transverse expansion and longitudinally boost-invariant
scaling flow are assumed \cite{dirk1,gersdorff,DumRi}. 
This approximation should be reasonable 
for central collisions at high energy, and around midrapidity. 
The model reproduces the measured $p_T$-spectra and rapidity densities
of a variety of hadrons at $\sqrt{s}=17.4A$~GeV (CERN-SPS energy),
when assuming the standard thermalization
(proper) time $\tau_i=1$~fm/c, and an entropy per net
baryon ratio of $s/\rho_B=45$ \cite{DumRi,hu_first,hu_main}.
Due to the higher density at midrapidity,
thermalization may be faster at BNL-RHIC energies -- here
we assume $\tau_i=0.6$~fm/c and $s/\rho_B=200$.
(With these initial conditions preliminary results on the multiplicity, the
transverse energy, the $p_T$-distribution of charged hadrons, and
the $\overline{p}/p$ ratio at $\sqrt{s}=130A$~GeV are described quite
well~\cite{DumRi,hu_first,hu_main}; below it will be shown that HBT
correlations of pions at small relative momenta do {\em not} depend
sensitively on these initial conditions.)
The energy density and baryon density are initially distributed in the
transverse plane according to a so-called ``wounded nucleon'' distribution
with transverse radius $R_T=6$~fm. For further details, we refer to
refs.~\cite{DumRi,hu_first,hu_main}.

We shall first discuss the radii of the $\pi^-\pi^-$ correlation functions
at hadronization. 
From the hydrodynamical solution in the forward light cone we determine
the hadronization hypersurface $\sigma^\mu$.
On that surface, the two-particle correlation function is given
by~\cite{schlei,dirk1}
\be \label{C2hadroniz}
C_2({\bf p_1},{\bf p_2}) = 1 + \frac{1}{{\cal N}}
   \left|
\int\d\sigma\cdot K e^{i\sigma\cdot q} f\left(u\cdot K/T\right)
   \right|^2
\ee
This assumes a chaotic (incoherent) and large source.
$f$ denotes a Bose distribution function.
The normalization factor ${\cal N}$ is given by the product of the
invariant single-inclusive distribution of $\pi^-$ evaluated at
momenta ${\bf p_1}$ and ${\bf p_2}$, respectively. $u^\mu$
denotes the four-velocity of the fluid on the   
hadronization surface $\sigma^\mu$, and $K^\mu=(p_1^\mu+p_2^\mu)/2$,
$q^\mu=p_1^\mu-p_2^\mu$ are the average four-momentum and the
relative four-momentum of the pion pair, respectively. 
For midrapidity pions $K_\parallel=q_\parallel=0$.
Note that eq.~(\ref{C2hadroniz}) accounts for the direct pions only
but not for decays of hadronic resonances (like $\rho^0\rightarrow   
\pi^+\pi^-$ etc.), which are known to affect the correlation
functions~\cite{schlei,reviews,wiedemannrep,Gyulassy:1989yr} and which
will be included below.

One usually employs a coordinate system in which
the {\it long} axis ($z$) is chosen parallel to
the beam axis, where the {\it out} direction
is defined to be parallel to the transverse momentum vector
${\bf K_T}=({\bf p_{1T}} + {\bf p_{2T}})/2$
of the pair, and the {\it side} direction is perpendicular
to both.

\begin{figure}[htp]
\centerline{\hspace{.4cm}\hbox{\epsfig
{figure=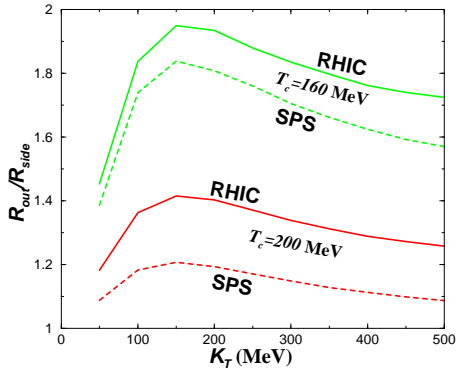,width=6cm}}}
\caption{Ratio of the inverse widths of the correlation function in
{\it out} and {\it side} direction at hadronization,
as a function of $K_T$; 
for $T_c\simeq160$~MeV and $\simeq200$~MeV, respectively.}
\label{hadroniz_Ro_Rs}
\end{figure}

From eq.(\ref{C2hadroniz}), the inverse widths of the correlation
function are obtained as $R_{\rm out} = \sqrt{\ln 2}/q_{\rm out}^*$ 
and $R_{\rm side}= \sqrt{\ln 2}/q_{\rm side}^*$
where $q_{\rm out}^*$, $q_{\rm side}^*$ are defined by 
$C_2(q_{\rm out}^*,q_{\rm side}=0)=
C_2(q_{\rm side}^*,q_{\rm out}=0)=3/2$. 
Due to the definition of the {\it out} and {\it side} direction, 
$R_{\rm out}$ probes the spatial {\it and} temporal extension 
of the source while $R_{\rm side}$ only probes the spatial extension. 
Thus the ratio $R_{\rm out}/R_{\rm side}$ gives a measure of the 
emission duration (see also eqs.(3)-(5) and discussion below). 
It has been suggested that the ratio $R_{\rm out}/
R_{\rm side}$ should increase strongly once the initial
entropy density $s_i$ becomes substantially larger than that of the hadronic 
gas at $T_c$~\cite{dirk1}.
Indeed, Fig.~\ref{hadroniz_Ro_Rs} shows that $R_{\rm out}/R_{\rm side}$
is much larger if $T_c$ is low, such that entropy conservation
dictates a long hadronization time.
The closer $T_c$ is to
the initial temperature $T_i (\approx300$~MeV for the BNL-RHIC initial
conditions), the faster $T_c$
is reached from above. For 1+1 dimensional isentropic
scaling expansion the time to complete hadronization is given by
$ \tau_H/\tau_i = {s_i}/s_H(T_c)$,
where $s_H(T_c)$ is the entropy density of the hadronic phase at $T_c$.
Of course, $s_H(T_c)$ increases with $T_c$ and so the
hadronization time decreases. As $R_{\rm out}$
grows with the duration of pion
emission~\cite{pratt86}, it must therefore decrease as $T_c$ increases.
Also, $R_{\rm out}/R_{\rm side}$ decreases towards large $K_T$ as
transverse flow reduces the scale of spatial homogeneity; that effect is
stronger for lower $T_c$ as transverse collective flow has more time to 
develop during the lifetime of the QGP.

We now proceed to include hadronic rescattering following hadronization. 
To describe the evolution of the hadrons towards
freeze-out obviously requires to go beyond perfect-fluid dynamics.
Here, we employ a semi-classical transport model that treats
each particular hadronic reaction channel (formation and decay of hadronic
resonance states and $2\rightarrow n$ scattering)
{\em explicitly}~\cite{bass98,bleicher99}. The transition 
at hadronization is performed by matching
the energy-momentum tensors and conserved currents of the
hydrodynamic solution and of the microscopic transport model, respectively
(for details, see~\cite{hu_main}). 
The microscopic model propagates each individual hadron
along a classical trajectory, and performs $2\rightarrow n$ and $1\rightarrow
m$ processes stochastically.
E.g., the total meson-meson cross section includes a 5~mb elastic
contribution as well as resonance excitation, which dominates the cross
section, via
\begin{eqnarray}  
\label{mbbreitwig}
\sigma^{M M'}_{\rm res}(\sqrt{s}) &=& \sum\limits_{R=M^*}
       \langle j_M, m_M, j_{M'}, m_{M'} \| J_R, M_R \rangle \nonumber \\
\times\hspace{1.5cm}&&\hspace{-1.7cm}\!\!\!\!\frac{2 S_R +1}{(2 S_M +1) 
(2 S_{M'} +1 )}\,
      \frac{\pi}{p^2_{\rm cms}}\,
        \frac{\Gamma_{R \rightarrow M M'} \Gamma_{\rm tot}}
             {(m_R - \sqrt{s})^2 + \Gamma_{\rm tot}^2/4}
\end{eqnarray}
with the total and partial $\sqrt{s}$-dependent decay widths
$\Gamma_{\rm tot}$ and $\Gamma_{R \rightarrow M M'}$. (Here, $\sqrt{s}$
refers to the cm-energy of the hadron-hadron scattering.)
The full decay width $\Gamma_{\rm tot}(m)$ of a resonance is
defined as the sum of all partial decay widths and depends on the
mass of the excited resonance. The sum extends over all resonance states
which have a decay channel into $M M'$.
The pole masses and partial decay widths are taken
from~\cite{PDG}.
The $\pi\pi$ cross section for example is dominated by the $\rho(770)$,
with additional contributions from higher energy states.
$\pi K$ scattering is either elastic or proceeds
through formation of mainly a $K^*(892)$ resonance.
Meson-baryon elastic and resonant scattering is also taken
into account. 
In this way, a good description of elastic and total pion cross sections
{\em in vacuum} is obtained~\cite{bleicher99}.
Collective medium-induced effects on the
pion scattering~\cite{mediumeff} are neglected at present
because {\em in a thermalized state} at $T= T_c$
there is less than one $\pi^-$ per phase space cell
$\d^3x \d^3p/(2\pi)^3$.

The distribution of freeze-out points of pions in the forward light-cone
is rather broad in time~\cite{husecond,microFO}.
Freeze-out occurs in a {\em four-dimensional} region within the
forward light-cone rather than on a three-dimensional ``hypersurface''.
The single-particle distribution of pions
is not altered very much during the dissipative hadronic phase because
hard collisions are rare~\cite{hu_first,hu_main}.
(The pressure $p$ is small and moreover $-p\d V$ mechanical work is
largely compensated by $+T\d S$ entropy production.)
There are, however, numerous soft collisions~\cite{husecond},
characterizing the dissipative evolution that approaches freeze out. 
That means that the hadronic system, when starting
from a state of local equilibrium at hadronization, disintegrates slowly
rather than emitting a ``flash'' of pions in a rapid decay.
This is fundamentally different from the ``explosive'' hadron production
from the decay of a classical background field at the confinement transition
via parametric resonance~\cite{Pisarski:2000eq}.

The solution of the microscopic transport provides the classical
phase space distribution of the hadrons at the points of their last
(strong) interaction. Bose-Einstein correlations are introduced
a posteriori by identifying the phase space distribution at freeze-out
with the Wigner density of the
source~\cite{wiedemannrep,Gyulassy:1989yr,Pratt:1990zq}, $S(x,K)$.
Corrections arise if the pions
undergo a stage of ``cascading'' from the space-time point of their
production to the point of their last interaction~\cite{padula90}.
However, from the above mentioned model for pion rescattering
we find that only $\sim 15-20\%$ of the pions in the
final state freeze out after an elastic scattering.
Rather, most pions emerge from the fragmentation of a
hadronic resonance, or are emitted directly from the hadronization
hypersurface. Thus the source can be considered chaotic.
A more detailed discussion will
be given elsewhere~\cite{to_come}.

We focus here on the so-called Gaussian radius parameters,
which are obtained from a saddle-point integration over 
$S(x,K)$~\cite{wiedemannrep,Chapman:1995nz}. The correlation
functions themselves, the value of the intercept, and the effects of
introducing finite momentum resolutions (as in the experimental analysis)
will be discussed in~\cite{to_come}.

The HBT-radius parameters characterizing the Gaussian {\sl ansatz} are
\begin{eqnarray}
\label{rs}
R_{\rm side}^2({\bf K_T})&=& \langle \tilde{y}^2 \rangle ({\bf K_T})\,,\\
R_{\rm out}^2({\bf K_T})&=& \langle (\tilde{x}-\beta_t \tilde{t})^2
\rangle ({\bf K_T}) = \langle\tilde{x}^2+\beta_t^2 \tilde{t}\,^2-2
 \beta_t\tilde{x}\tilde{t}\rangle\,,\label{ro}\\
R_{\rm long}^2({\bf K_T})&=& \langle (\tilde{z}-\beta_l \tilde{t})^2
\rangle ({\bf K_T})\,,\label{rl}
\end{eqnarray}
with
$\tilde{x}^{\mu}({\bf K_T}) = x^{\mu} - \langle {x}^{\mu}\rangle({\bf K_T})$ 
being the space-time coordinates relative 
to the momentum dependent {\it effective source centers}.
The average in (\ref{rs})-(\ref{rl}) is taken over the
emission function, i.e.
$\langle f \rangle(K)= \int d^4x f(x) S(x,K) / \int d^4x S(x,K)$.
In the {\it osl} system ${\bf \beta}=(\beta_t,0,\beta_l)$, where
${\bf \beta}={\bf K}/E_K$ and $E_K=\sqrt{m^2+{\bf K}^2}$.
Below, we shall cut on midrapidity pions ($\beta_l\sim0$), thus the radii
are obtained in the {\it longitudinally comoving frame}.
In the absence of $\tilde{x}$-$\tilde{t}$ correlations,
a large duration of emission $\Delta \tau = \surd
{\langle \tilde{t}\,^2 \rangle}$ increases $R_{\rm out}$ relative
to $R_{\rm side}$~\cite{pratt86,schlei,dirk1}.
\begin{figure}[htp]
\centerline{\hspace{.4cm}\hbox{\epsfig{figure=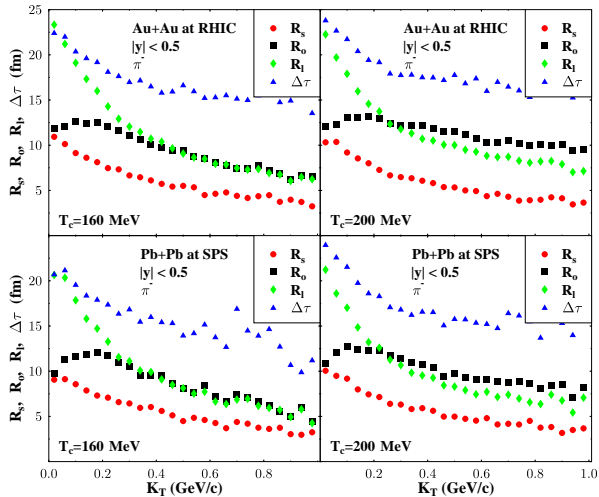,width=8cm}}}
\caption{HBT radii $R_{\rm out},\,R_{\rm side},\,
R_{\rm long},$ and emission duration $\Delta \tau$ at freeze-out
as a function of $K_T$;
for central collisions at RHIC (above) and 
SPS (below) and $T_c\simeq160\,$MeV (left) and $T_c\simeq200\,$MeV (right).}
\label{fig_rosl}
\end{figure}
Fig.~\ref{fig_rosl} shows the HBT radii and $\Delta \tau$.
Also, we have varied the bag parameter $B$ from $380$~MeV/fm$^3$ to
$720$~MeV/fm$^3$, corresponding to critical temperatures of
$T_c\simeq160$~MeV and $T_c\simeq200$~MeV,
respectively. Note that within the bag model this automatically corresponds
to a large variation of the latent heat $\sim4B$ as well,  
serving  the purpose of proving the (in)sensitivity of the HBT radii
to variations of the phase transition parameters in this model.
We also note that
changing $T_c$ implies variation of the longitudinal and transverse flow
profile on the hadronization hypersurface (which is the initial condition for
the subsequent hadronic rescattering stage) over a broad range.

It is obvious from Fig.~\ref{fig_rosl} that the
HBT radii are rather similar for all cases considered.
They depend only weakly on the specific entropy, 
on the critical temperature and latent heat for the transition, 
on the thermalization time $\tau_i$, or on the initial condition for the
hadronic rescattering stage.
Thus, the properties of the QGP-phase are not directly reflected
in the HBT radii, which are essentially determined by the
large space-time volume of the hadronic rescattering stage. 
Soft hadronic scattering occurs over a long time-span after
hadronization~\cite{husecond}, and therefore
the pion mean free path increases gradually towards
freeze-out~\cite{Prakash:1993bt}. Pions are
emitted over a broad time interval and from the entire volume. Hence,
the emission duration is large, see Fig.~\ref{fig_rosl}.

\begin{figure}[htp]
\centerline{\hspace{.8cm}\hbox{
\epsfig{figure=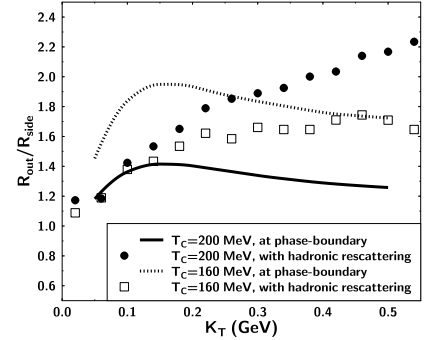,width=6cm}}}
\caption{$R_{\rm out}/R_{\rm side}$ for RHIC initial conditions,
as a function of $K_T$ at freeze-out
(symbols) and at hadronization (lines).}\label{osratio}
\end{figure}
In Fig.~\ref{osratio} we compare $R_{\rm out}/R_{\rm side}$ at hadronization
to that at freeze out. Clearly, up to $K_T\sim200\,$MeV
$R_{\rm out}/R_{\rm side}$ is independent of $T_c$, or $s_H(T_c)$,
if hadronic rescatterings are taken into account. 
Moreover, at higher $K_T$ the dependence on $T_c$ is even reversed: 
for high $T_c$ the $R_{\rm out}/R_{\rm side}$ ratio even exceeds that
for low $T_c$.
Higher $T_c$ speeds up hadronization but on the other hand prolongs 
the dissipative hadronic phase that dominates the HBT radii. This is because
during the non-ideal hadronic expansion the scale of spatial homogeneity
of the pion density distribution increases, as the pions fly away from the
center, but the transverse flow can hardly increase to counteract
(similar to a Hubble expansion at small pressure). 
Therefore, after hadronic rescattering 
$R_{\rm out}/R_{\rm side}$ does not drop towards
higher $K_T$ (in the range $K_T\lton 3m_\pi$).

Experimental data~\cite{App98:epj_expansion,blume01} for central Pb+Pb 
collisions at
$\sqrt{s}=17.4A$~GeV indicate smaller HBT radii than seen in
Fig.~\ref{fig_rosl}, in particular at $K_T\le100$~MeV. All three radii are
$\le10$~fm. One should keep in mind though that experimental uncertainties
are largest at small $K_T$, where also Coulomb, acceptance and 
efficiency  corrections are largest and experimental
momentum resolution effects are crucial. 
At $K_T >200$~MeV, the calculated HBT radii 
agree better with data. 
The experimental momentum resolution or the fit procedures
could in effect reduce the extracted radii --
these rather technical issues will be investigated
in a detailed forthcoming study~\cite{to_come}. 
On the other hand, our calculation gives a similar
$K_T$-dependence of the three radii 
as seen in the SPS data: strong decrease 
of $R_{\rm long}$, weaker decrease of $R_{\rm side}$ and rather flat 
behavior of $R_{\rm out}$ with $K_T$. Most striking is 
that a new preliminary NA49 data analysis~\cite{blume01} shows a rising 
$R_{\rm out}/R_{\rm side}$ ratio, rather similar to our results
for SPS energy with $T_c\simeq160\,$MeV~\cite{to_come}.
 
Nevertheless, good agreement with the data might be not possible to
achieve for $R_{\rm out}$ and $R_{\rm long}$. That would point at
too large duration of emission, $\Delta\tau$, and too large spatial homogeneity
scale in this model scenario of a first order phase transition 
from a thermalized QGP to a hadron gas phase. 
Note that single-particle spectra above $p_T\sim m_\pi$
are described quite well~\cite{DumRi,hu_first,hu_main}. This
emphasizes the importance of interferometry as a crucial test of the QGP model
with hadronization through a mixed phase,
as HBT probes the small momentum transfer dynamics and the spatial homogeneity
scale at freeze out.

At the higher RHIC energy, the model can be tested at even higher
$K_T$, where experimental resolutions and corrections will be less of an issue.
Also, higher energy densities will be achieved, such that the QGP model
should in principle be more reliable. For central collisions of Au nuclei at
$\sqrt{s}=130A$~GeV, preliminary data from STAR gives
$R_{\rm out}/R_{\rm side}\simeq1.1$ at small $K_T$~\cite{Lisa00}.
Results from RHIC at higher $K_T$ will test whether a long-lived
{\em hadronic} soft-rescattering stage, associated with the formation and
hadronization of an equilibrated QGP state is indeed seen in
heavy ion collisions at the highest presently attainable energies.
\vspace*{-0.4cm}
\acknowledgements 
\vspace*{-0.4cm}
We are grateful to M.\ Gyulassy, M.\ Lisa, L.\ McLerran, 
S.\ Panitkin, S.\ Pratt, D.H.\ Rischke, H. St\"ocker, U.A.\ Wiedemann,
and N.\ Xu for many valuable comments. We thank the
UrQMD collaboration for permission to use the UrQMD transport model.
S.S.\ has been supported in part by the Humboldt Foundation
through a Feodor Lynen Fellowship and DOE grant 
DE-AC02-98CH10886. 
S.A.B.\ acknowledges support from DOE grant DE-FG02-96ER40945, 
and A.D.\ from DOE grant DE-FG-02-93ER-40764.

\end{document}